\documentclass[11pt]{iopart}

\usepackage{graphicx}
\usepackage{amssymb}
\usepackage{iopams}
\usepackage{amsthm}
\usepackage{latexsym}
\usepackage{pst-all}
\usepackage{capt-of}
\usepackage{epsfig}

\newcommand{\E}{{\cal{E}}}

\newcommand{\I}{{\rm i}}
\renewcommand{\a}{\alpha}
\renewcommand{\d}{{\rm d}}

\newcommand{\be}{\begin{equation}}
\newcommand{\ee}{\end{equation}}
\newcommand{\bea}{\begin{eqnarray}}
\newcommand{\eea}{\end{eqnarray}}

\newcommand{\nabv}{\mbox{\boldmath$\nabla$}}

\def\J#1#2#3#4{#1 {\it #2} {\bf #3} #4}
\def\PTP{\it Prog. Theor. Phys.}

\def\PRD{{\it Phys. Rev.} D}
\def\PR{\it Phys. Rev.}

\def\CQG{\it Class. Quantum Grav.}
\def\ZP{\it Z. Phys.}
\def\PLA{\it Phys. Lett. A}

\begin{document}

\title[Poynting vector reexamined]
{Poynting vector in stationary axisymmetric electrovacuum
spacetimes reexamined}

\author{V S Manko$^1$, E D Rodchenko$^2$, B I Sadovnikov$^2$
and\\ J Sod--Hoffs$^1$}

\address{$^1$ Departamento de F\'{i}sica, Centro
de Investigaci\'{o}n y de Estudios Avanzados del IPN, A.P. 14-740,
07000 M\'{e}xico D.F., Mexico}
\address{$^2$ Department of Quantum Statistics and Field Theory,
Lomonosov Moscow State University, Moscow 119899, Russia}

%%% Abstract
\begin{abstract}
A simple formula, invariant under the duality rotation $\Phi\to
{\rm e}^{i\a}\Phi$, is obtained for the Poynting vector within the
framework of the Ernst formalism, and its application to the known
exact solutions for a charged massive magnetic dipole is
considered.
\end{abstract}

 %\submitto{\CQG}
\maketitle

%%%%%%%%%%%%%%%%%%%%%%%%%%%%%%%%%%%%%%%%%%%%%%%%%%%%%%%%%%%%%%%%%%
\section{Introduction}
%%%%%%%%%%%%%%%%%%%%%%%%%%%%%%%%%%%%%%%%%%%%%%%%%%%%%%%%%%%%%%%%%%

In a recent paper \cite{HGP} the vorticity tensor and Poynting
vector have been analyzed in the context of Bonnor's
frame--dragging effect \cite{Bon} occurring in the field of a
static mass endowed with electric charge and magnetic dipole
moment. In spite of presenting several useful general formulae,
the authors of \cite{HGP}, in our opinion, failed to achieve one
of the main goals of their research which was the demonstration in
the explicit form, using a particular exact solution, the
appearance of the predicted by Bonnor factor $qb$, $q$ being the
electric charge and $b$ the magnetic dipole moment, in the
expression for the component $\varphi$ of Poynting vector. This
may have the following two explanations: firstly, the formula (28)
of \cite{HGP} defining $S^\varphi$ is not quite appropriate for
concrete applications, even in the simplest cases such as Manko's
solution \cite{Man} for a magnetized Kerr--Newman mass utilized in
\cite{HGP}. And, secondly, the determinantal form of writing
specific exact solutions employed in \cite{HGP} only permits to
arrive at some formal expressions which need to be further
simplified by expanding the determinants.

A detailed examination of the paper \cite{HGP} reveals that it
contains an intrinsic inconsistency which is the attribution to
the same function --- the electric component of the
electromagnetic 4--potential --- different signs in the field
equations and during the calculation of the Poynting vector. To
make things worse, it appears that the determinantal expressions
of \cite{HGP} which are entirely taken over from the paper
\cite{RMM} devoted to the multisoliton electrovac solution, are
{\it all} presented with errors, including the definitions of the
quantities $h_l(\a_n)$. Such distorted formulae can be neither
reproduced nor used in any physical analysis; neither they can be
considered as a substitute to the elegant original formulae
defining Manko's electrovac solution \cite{Man}.

The objective of the present paper is twofold: ($i$) the
derivation of a simple formula for the component $\varphi$ of the
Poynting vector which would be consistent with the Ernst formalism
and based on it, and ($ii$) its subsequent application to two
exact solutions for a charged massive magnetic dipole for
providing an explicit demonstration of Bonnor's frame--dragging
effect via the Poynting vector. The point ($i$) of our objective
will be achieved by first reviewing the Ernst formalism of complex
potentials, paying attention to the historical notations used in
the original Ernst's article \cite{Ern}, and then deriving the
expression for $S^\varphi$ in which the simplifications will be
possible thanks to the use of the differential relations provided
by the Ernst formalism. The main advantage of the new formula for
$S^\varphi$ thus obtained will consist in the absence in it of the
metric function $\omega$ which is normally the most complicated
metric coefficient in the electrovac spacetimes. The formula will
also permit us to draw an important conclusion about the
invariance of $S^\varphi$ under the duality rotation $\Phi\to {\rm
e}^{i\a}\Phi$, $\a={\rm const}$. The point ($ii$) will be realized
on the way of applying the formula for $S^\varphi$ to Manko's
electrovac solution \cite{Man}, and to the rational function
solution constructed in the paper \cite{MSM}. Since the latter
solution is written in the spheroidal coordinates $(x,y)$, a
corresponding formula for $S^\varphi$ in these coordinates will be
also worked out.

%%%%%%%%%%%%%%%%%%%%%%%%%%%%%%%%%%%%%%%%%%%%%%%%%%%%%%%%%%%%%%
\section{Brief review of the Ernst formalism}
%%%%%%%%%%%%%%%%%%%%%%%%%%%%%%%%%%%%%%%%%%%%%%%%%%%%%%%%%%%%%%

In his famous paper \cite{Ern} Ernst reduced the problem of
finding stationary axisymmetric electrovacuum solutions of the
Einstein--Maxwell equations to solving a concise system of two
differential equations for the complex potentials $\E$ and $\Phi$:
\bea\label{EE} ({\rm Re}\E+\Phi\bar\Phi)\Delta\E=
(\nabv\E+2\bar\Phi\nabv\Phi)\nabv\E, \nonumber\\ ({\rm
Re}\E+\Phi\bar\Phi)\Delta\Phi=
(\nabv\E+2\bar\Phi\nabv\Phi)\nabv\Phi, \eea where a bar over a
symbol denotes complex conjugation, $\Delta$ and $\nabv$ are the
three--dimensional Laplacian and gradient operators, respectively
(in the Weyl--Papapetrou cylindrical coordinates ($\rho,z$) \be
\Delta A:=A_{,\rho,\rho}+\rho^{-1}A_{,\rho}+A_{,z,z}, \quad \nabv
A\cdot \nabv B:=A_{,\rho}B_{,\rho}+A_{,z}B_{,z}.) \ee

The relation of the potentials $\E$ and $\Phi$ to the coefficients
in the stationary axisymmetric line element
\begin{equation}\label{papa}
\d s^2=g_{ik}\d x^i\d x^k= f^{-1}[e^{2\gamma} (\d\rho^2 + \d z^2)
+ \rho^2 \d\varphi^2]- f( \d t - \omega \d\varphi)^2
\end{equation}
($f$, $\omega$ and $\gamma$ are functions of $\rho$ and $z$ only;
$x^1=\rho$, $x^2=z$, $x^3=\varphi$, $x^4=t$) and to the $\varphi$
and $t$ components of the electromagnetic 4--potential \be
A_i=(0,0,A_3,-A_4) \label{4-pot} \ee is defined by the following
equations: \bea &&\E=f-\Phi\bar\Phi+\I\chi, \quad \Phi=A_4+\I
A_3^{'}, \label{EF} \\ &&\chi_{,\rho}=\rho^{-1}f^2\omega_{,z}
-2{\rm Im}(\bar\Phi\Phi_{,\rho}), \quad
\chi_{,z}=-\rho^{-1}f^2\omega_{,\rho} -2{\rm
Im}(\bar\Phi\Phi_{,z}), \label{chi}  \\
&&A_{3,\rho}^{'}=\rho^{-1}f(A_{3,z}-\omega A_{4,z}), \quad
A_{3,z}^{'}=-\rho^{-1}f(A_{3,\rho}-\omega A_{4,\rho}), \label{A3'}
\eea and these do not involve the metric function $\gamma$ which
can be found, for known $\E$ and $\Phi$, from the system of the
first order differential equations \bea
\gamma_{,\rho}&=&\frac{1}{4} \rho
f^{-2}[(\E_{,\rho}+2\bar\Phi\Phi_{,\rho})
(\bar\E_{,\rho}+2\Phi\bar\Phi_{,\rho}) \nonumber\\
&&-(\E_{,z}+2\bar\Phi\Phi_{,z}) (\bar\E_{,z}+2\Phi\bar\Phi_{,z})]
-\rho f^{-1}(\Phi_{,\rho}\bar\Phi_{,\rho}-
\Phi_{,z}\bar\Phi_{,z}), \nonumber\\
\gamma_{,z}&=&\frac{1}{2} \rho f^{-2}{\rm
Re}[(\E_{,\rho}+2\bar\Phi\Phi_{,\rho})
(\bar\E_{,z}+2\Phi\bar\Phi_{,z})] -2\rho f^{-1}{\rm
Re}(\bar\Phi_{,\rho}\Phi_{,z}), \label{eq_g} \eea the
integrability condition of which is the system (\ref{EE}).

Once the Ernst equations (\ref{EE}) are solved, the metric
function $f$, the electric potential $A_4$ and the functions
$\chi$ and $A_3^{'}$ can be found from the formulae \be f={\rm
Re}(\E)+\Phi\bar\Phi, \quad \chi={\rm Im}(\E), \quad A_4={\rm
Re}(\Phi), \quad A_3^{'}={\rm Im}(\Phi), \label{fA4} \ee which, in
turn, permit to obtain the metric coefficient $\omega$ by
integrating equations (\ref{chi}), and subsequently the magnetic
potential $A_3$ by integrating equations (\ref{A3'}). Lastly, the
function $\gamma$ is obtainable from the system (\ref{eq_g}).

Therefore, the knowledge of the Ernst potentials $\E$ and $\Phi$
is sufficient for the reconstruction of the whole metric and of
the electromagnetic field outside the sources.

It is worth pointing out that historically the sign `--' in the
formula (\ref{4-pot}) was chosen for convenience, with the idea to
avoid the minus sign in the formula (\ref{EF}) defining the
potential $\Phi$. It is the conventional character of this choice
which later allowed some researchers to formally redefine at their
own taste the $t$--component of the potential $A_i$, so that the
Ernst potential $\Phi$ may be written for instance in the form
\cite{Tom} \be \Phi=-A_t+\I A_\varphi^{'}, \quad A_t:=-A_4, \quad
A_\varphi^{'}:=A_3^{'}. \label{At} \ee

The only thing one should be aware of if one wants to redefine the
original Ernst's electric potential $A_4$, is not to forget to
carry out the respective changes in equations (\ref{EF}) and
(\ref{A3'}). The paper \cite{HGP} is very instructive in this
connection. Its authors first wrote out (up to a couple misprints)
the original Ernst's formulae, but then, probably not knowing or
having forgotten  formula (\ref{4-pot}), they carried out the
calculation of the components of the electromagnetic
energy--momentum tensor as if they were working not with the
potential $A_4$ but with $-A_4$. Such inconsistency did not permit
them to take any advantage of the Ernst formalism and carry out
further simplifications of the formula for $S^\varphi$.

%%%%%%%%%%%%%%%%%%%%%%%%%%%%%%%
\section{Poynting vector in the Ernst formalism}
%%%%%%%%%%%%%%%%%%%%%%%%%%%%%%%

The Poynting vector $S^\a$ (in this paper the Greek indices take
the values 1, 2 and 3) which was considered in application to the
stationary axisymmetric electrovacuum case in \cite{HGP} is
defined by the formula\footnote{In comparison with the paper
\cite{HGP} we have changed the sign in the definition of $T_{ik}$
following the book \cite{MTW}, and consequently changed the sign
in the original formula for $S^\a$ used by Herrera et al.} \be
S^\a=T^{\a i}u_i, \label{Poynt} \ee where $T^{ik}$ is the
energy--momentum tensor of the electromagnetic field \be
T_{ik}=\frac{1}{4\pi}(F_{il}F_k{}^l-\frac{1}{4}g_{ik}F_{lm}F^{lm}),
\label{Tik} \ee $F_{ik}$ being the electromagnetic field tensor
\be F_{ik}=A_{k,i}-A_{i,k}, \label{Fik} \ee and $u^i$ is the
4--velocity vector with the components \be u^i=(0,0,0,f^{-1/2}),
\quad u_i=(0,0,f^{1/2}\omega,-f^{1/2}), \quad u^i u_i=-1.
\label{ui} \ee

Taking into account that \bea g^{11}=g^{22}=f{\rm e}^{-2\gamma},
\,\, g^{33}=\rho^{-2} f, \,\, g^{34}=\rho^{-2} f\omega, \,\,
g^{44}=-f^{-1}+\rho^{-2} f\omega^2, \nonumber\\ F_{13}=A_{3,\rho},
\quad F_{14}=-A_{4,\rho}, \quad F_{23}=A_{3,z}, \quad
F_{24}=-A_{4,z}, \nonumber\\ T_{11}=\frac{1}{4\pi}[\rho^{-2}
fA_{3,\rho}^2+(\rho^{-2} f\omega^2-f^{-1})A_{4,\rho}^2 -2\rho^{-2}
f\omega A_{3,\rho}A_{4,\rho}-\frac{1}{2}A], \nonumber\\
T_{12}=\frac{1}{4\pi}[\rho^{-2} f(A_{3,\rho}-\omega
A_{4,\rho})(A_{3,z}-\omega A_{4,z})-f^{-1}A_{4,\rho}A_{4,z}, \nonumber\\
T_{22}=\frac{1}{4\pi}[\rho^{-2} fA_{3,z}^2+(\rho^{-2}
f\omega^2-f^{-1})A_{4,z}^2 -2\rho^{-2} f\omega
A_{3,z}A_{4,z}-\frac{1}{2}A], \nonumber\\
T_{33}=\frac{1}{4\pi}f{\rm e}^{-2\gamma}[(\nabv A_3)^2
-\frac{1}{2}(\rho^2 f^{-1}-f\omega^2)A], \nonumber\\
T_{34}=-\frac{1}{4\pi}f{\rm e}^{-2\gamma}(\nabv A_3\nabv A_4)
+\frac{1}{2}f\omega A, \nonumber\\ T_{44}=\frac{1}{4\pi}f{\rm
e}^{-2\gamma}[(\nabv A_4)^2 +\frac{1}{2}f A], \nonumber\\
A:=\rho^{-2} f(\nabv A_3)^2+(\rho^{-2} f\omega^2-f^{-1})(\nabv
A_4)^2-2\rho^{-2} f\omega \nabv A_3\nabv A_4, \eea it is
straightforward  to verify that $S^\rho=S^z=0$ identically, while
the calculation of the non--zero component $S^\varphi$ leads to
the expression \be S^\varphi=\frac{f^{3/2}{\rm
e}^{-2\gamma}}{4\pi\rho^2}[\omega(\nabv A_4)^2-\nabv A_3\nabv
A_4]. \label{Sf} \ee

Formula (\ref{Sf}) differs from the formula (28) of \cite{HGP} in
the sign of the second term in brackets, and it also contains the
factor 4 in the denominator on the right--hand side which is
missing in \cite{HGP}. The difference in sign is explained by the
already mentioned inconsistency in the use of the $t$--component
of the potential $A_i$ taking place in \cite{HGP}.

Formula (\ref{Sf}) is congruent with the equations (\ref{EF}) and
(\ref{A3'}) and, as a result, can be further simplified. Indeed,
by passing in (\ref{Sf}) from $A_3$ to $A_3^{'}$ with the aid of
equations (\ref{A3'}), we obtain \be S^\varphi=\frac{\sqrt{f}{\rm
e}^{-2\gamma}}{4\pi\rho}(A_{4,\rho} A_{3,z}^{'}-A_{4,z}
A_{3,\rho}^{'}), \label{Sf2} \ee thus getting rid of the metric
function $\omega$. Then, making use of (\ref{EF}), we arrive at
the final result \be S^\varphi=\frac{\sqrt{f}{\rm
e}^{-2\gamma}}{4\pi\rho}{\rm Im}(\bar\Phi_{,\rho}\Phi_{,z}).
\label{Sf3} \ee

Formula (\ref{Sf3}) is by far more suitable for the use in
concrete applications than formula (\ref{Sf}) since in order to
see whether or not $S^\varphi$ is zero, the knowledge of only one
Ernst's potential $\Phi$ is sufficient. From (\ref{Sf3}) also
follows that $S^\varphi$ is invariant under the duality rotation
transformation \be \Phi\to {\rm e}^{\I\a}\Phi, \quad \a={\rm
const} \label{dual} \ee which means in particular that $S^\varphi$
will be equal to zero (and consequently no frame--dragging will
occur) in the electrovacuum spacetimes representing a static mass
endowed with both electric and magnetic dipole moments if the
corresponding exact solution was obtained from the magnetostatic
(or electrostatic) solution by means of a duality rotation. More
generally, $S^\varphi$ of some electrovac solution will be equal
to zero if the potential $\E$ of that solution is a real function,
and its potential $\Phi$ can be made a real or pure imaginary
function via exclusively an appropriate duality rotation.

We shall now illustrate the use of formula (\ref{Sf3}) with two
examples.

%%%%%%%%%%%%%%%%%%%%%%
\section{Examples}
%%%%%%%%%%%%%%%%%%%%%%

As the first example we will consider Manko's solution \cite{Man}
for a charged, magnetized, rotating mass, restricting ourselves to
the case when the total angular momentum is equal to zero
(together with the whole set of the rotational multipole moments).
The Ernst potentials $\E$ and $\Phi$ of this solution have the
form \cite{Man} \bea \E&=&\frac{A-B}{A+B}, \quad
\Phi=\frac{C}{A+B}, \nonumber\\
A&=&\kappa_+^2[(m^2-q^2-b)(R_+r_-+R_-r_+)+\I q\kappa_-
(R_+r_--R_-r_+)] \nonumber\\
&+&\kappa_-^2[(m^2-q^2+b)(R_+r_++R_-r_-)-\I q\kappa_+
(R_+r_+-R_-r_-)] \nonumber\\ &-&4b^2(R_+R_-+r_+r_-), \nonumber\\
B&=&m\kappa_+\kappa_-\{\kappa_+\kappa_-(R_++R_-+r_++r_-)
-(m^2-q^2)(R_++R_- \nonumber\\ &-&r_+-r_-)+\I
q[(\kappa_+-\kappa_-)(R_+-R_-)-(\kappa_++\kappa_-)(r_+-r_-)]\},
\nonumber\\
C&=&\kappa_+\kappa_-\{q\kappa_+\kappa_-(R_++R_-+r_++r_-)
-q(m^2-q^2) \nonumber\\ &\times&(R_++R_--r_+-r_-)+\I
[\kappa_+(q^2+b)(R_+-R_--r_++r_-) \nonumber\\
&-&\kappa_-(q^2-b)(R_+-R_-+r_+-r_-)]\}, \nonumber\\
R_\pm&=&\sqrt{\rho^2+[z\pm\frac{1}{2}(\kappa_++\kappa_-)]^2},
\quad
r_\pm=\sqrt{\rho^2+[z\pm\frac{1}{2}(\kappa_+-\kappa_-)]^2}, \nonumber\\
\kappa_\pm&=&\sqrt{m^2-q^2\pm 2b}, \label{EF_man} \eea where the
arbitrary real parameters $m$, $q$ and $b$ are, respectively, the
total mass, total charge and magnetic dipole moment of the source.
On the upper part of the symmetry axis ($\rho=0$,
$z>(\kappa_++\kappa_-)/2$) the potentials (\ref{EF_man}) take the
form \be \E(\rho=0,z)=\frac{z-m}{z+m}, \quad
\Phi(\rho=0,z)=\frac{qz+\I b}{z(z+m)}, \label{axis_man} \ee and
these axis data are sufficient for the construction of $\E$ and
$\Phi$ in the whole space (i.e., for arriving at formulae
(\ref{EF_man})) with the aid of Sibgatullin's method \cite{Sib}.
The metric functions $f$ and $\gamma$ which enter the expression
for $S^\varphi$ are defined for the Manko solution in terms of
$A$, $B$ and $C$ by the formulae \be f=\frac{A\bar A-B\bar B+C\bar
C} {(A+B)(\bar A+\bar B)}, \quad {\rm e}^{-2\gamma}=
\frac{16\kappa_+^4\kappa_-^4 R_+R_-r_+r_-}{A\bar A-B\bar B+C\bar
C}. \label{fg_man} \ee Since the metric coefficient $\omega$ which
has a rather complicated form is not needed for the calculation of
$S^\varphi$, it is not given here.

In the absence of one of the parameters $q$ or $b$ the potential
$\E$ becomes a real function, while $\Phi$ becomes a real ($b=0$)
or pure imaginary ($q=0$) function. During the reduction to the
magnetostatic case the disappearance of the imaginary part of $\E$
is obvious because of the presence of the factor $\I q$ in $A$ and
$B$. At the same time, in the electrostatic limit ($b=0$) the
vanishing of imaginary quantities and the reduction to the
well--known Reissner--Nordstr\"om solution is not that trivial,
and we find it instructive to show how this limit can be performed
(we observe that in \cite{HGP} the electrostatic limit of Manko's
solution was accomplished in an absolutely erroneous way, see
formulae (39) of \cite{HGP} where each term is artificially
multiplied by $b$ for getting the factor $\I qb$). Setting $b=0$
in (\ref{EF_man}) leads to \be \kappa_+=\kappa_-=\sqrt{m^2-q^2},
\quad r_+=r_-=\sqrt{\rho^2+z^2}, \label{estatic} \ee and then it
is easy to see that the sum of the imaginary terms in the first
and second lines of the expression for $A$ will be equal to zero
and, besides, the imaginary terms in $B$ will cancel out because
$r_--r_+=0$ and $\kappa_+-\kappa_-=0$. In the analogous way $C$
also becomes a real function in the limit $b=0$.

The calculation of the component $S^\varphi$ of the Poynting
vector with the aid of formulae (\ref{Sf3}), (\ref{EF_man}) and
(\ref{fg_man}) is straightforward and does not exhibit any
difficulty; the resulting expression is \be S^\varphi=\frac{128\,
q\, b\,\kappa_+^6\kappa_-^6F(\rho,z,R_\pm,r_\pm)}
{\pi(A+B)^{5/2}(\bar A+\bar B)^{5/2}(A\bar A-B\bar B+C\bar
C)^{1/2}}, \ee where $F(\rho,z,R_\pm,r_\pm)$ is some coefficient
which is not written down explicitly here because of its
cumbersome form. The appearance of the factor $qb$ in the
numerator of $S^\varphi$ is the desired result at which the above
calculation was aimed. It clearly demonstrates that the
non--vanishing of the Poynting vector in the solution for a static
mass possessing an electric charge and magnetic dipole moment is
due to the coexistence of the electric and magnetic fields.

Another solution appropriate for the description of the exterior
field of a static mass endowed with both the electric charge and
magnetic dipole moment is the three--parameter specialization of
the electrovac solution constructed in the paper \cite{MSM}. Its
characteristic feature is that it admits a rational functions
representation in the ellipsoidal coordinates. The Ernst
potentials $\E$ and $\Phi$ of the MSM solution have the form \bea
\E&=&\frac{A-B}{A+B}, \quad \Phi=\frac{C}{A+B}, \nonumber\\
A&=&2[(\kappa^2x^2-\delta y^2)^2-d^2]-2\I\kappa q b xy(1-y^2),
\nonumber\\ B&=&m[2\kappa^3x(x^2-1)+(1-y^2)(2\kappa\delta x-\I
qby)], \nonumber\\ C&=&2\kappa^2(x^2-1)(\kappa qx+\I by)
+(1-y^2)[2\kappa q\delta x-\I by(q^2-2\delta)], \label{EF_msm}
\eea where \bea x=\frac{1}{2\kappa}(r_++r_-), \quad
y=\frac{1}{2\kappa}(r_+-r_-), \quad
r_\pm=\sqrt{\rho^2+(z\pm\kappa)^2}, \nonumber\\
\kappa=\sqrt{d+\delta}, \quad d=\frac{1}{4}(m^2-q^2), \quad
\delta=\frac{b^2}{m^2-q^2}, \label{kappa} \eea the interpretation
of the arbitrary real parameters $m$, $q$ and $b$ being exactly
the same as in the previous example, i.e., mass, charge and
magnetic dipole moment of the source, respectively.

Both the magnetostatic and electrostatic limits of the solution
(\ref{EF_msm}) are classical: in the absence of the electric
charge it reduces to Bonnor's solution for a massive magnetic
dipole \cite{Bon2}, while in the absence of the magnetic field it
represents the charged Darmois solution \cite{Ern2}.

In the spheroidal coordinates ($x,y$) the line element
(\ref{papa}) assumes the form \bea \d s^2&=&\kappa^2f^{-1}\Bigl[
{\rm e}^{2\gamma}(x^2-y^2)\Bigl(\frac{\d x^2}{x^2-1}+ \frac{\d
y^2}{1-y^2}\Bigr)+(x^2-1)(1-y^2)\d\varphi^2\Bigr] \nonumber \\
&&-f(\d t-\omega\d\varphi)^2, \label{papa_xy} \eea and the metric
functions $f$, $\gamma$ and $\omega$ of the MSM solution are
defined by the following expressions: \bea f&=&E/D, \quad {\rm
e}^{-2\gamma}=16\kappa^8(x^2-y^2)^4/E, \quad \omega=qb(1-y^2)F/E,
\nonumber\\
E&=&4[\kappa^2(x^2-1)+\delta(1-y^2)]^4-4\kappa^2q^2b^2y^4
(x^2-1)(1-y^2), \nonumber\\ D&=&4[(\kappa^2x^2-\delta
y^2)^2+\kappa^3mx(x^2-1)+\kappa m\delta x(1-y^2)-d^2]^2
\nonumber\\ &&+q^2b^2y^2(2\kappa x+m)^2(1-y^2)^2, \nonumber\\
F&=&[\kappa^2(x^2-1)+\delta(1-y^2)]^2\{4[\kappa^2(x^2-1)
+\delta(1-y^2)]+(1-y^2) \nonumber\\ &&\times(4\kappa m
x+2m^2-q^2)\}+2\kappa^2y^2(x^2-1) \nonumber\\
&&\times\{\kappa m x[(2\kappa x+m)^2 -4\delta
y^2-q^2]-2\kappa^2q^2x^2-8d\delta y^2\}. \label{metric_msm} \eea

For the calculation of $S^\varphi$ in the spheroidal coordinates
($x,y$) it is necessary to carry out an appropriate coordinate
change in (\ref{Sf2}) and (\ref{Sf3}). The formulae relating
$\rho$ and $z$ to $x$ and $y$ are \be
\rho=\kappa\sqrt{(x^2-1)(1-y^2)}, \quad z=\kappa xy, \label{rzxy}
\ee and the partial derivatives with respect to $\rho$ and $z$ can
be changed to the derivatives with respect to $x$ and $y$ with the
aid of the formulae \bea \frac{\partial}{\partial\rho}&=&
\frac{\sqrt{(x^2-1)(1-y^2)}}{\kappa(x^2-y^2)}
\Bigl(x\frac{\partial}{\partial x}-y\frac{\partial}{\partial
y}\Bigr), \nonumber \\ \frac{\partial}{\partial z}&=&
\frac{y(x^2-1)}{\kappa(x^2-y^2)}\cdot\frac{\partial}{\partial x}
+\frac{x(1-y^2)}{\kappa(x^2-y^2)}\cdot\frac{\partial}{\partial y}.
\label{deriv_xy} \eea

Then for the terms containing derivatives of the potentials $A_4$
and $A_3^{'}$ with respect to $\rho$ and $z$ we get the expression
\be A_{4,\rho}A_{3,z}^{'}-A_{4,z}A_{3,\rho}^{'}=
\frac{\sqrt{(x^2-1)(1-y^2)}}{\kappa^2(x^2-y^2)}
(A_{4,x}A_{3,y}^{'}-A_{4,y}A_{3,x}^{'}), \label{coef_xy} \ee so
that formula (\ref{Sf3}) in the coordinates ($x,y$) finally
rewrites as \be S^\varphi=\frac{\sqrt{f}{\rm
e}^{-2\gamma}}{4\pi\kappa^3(x^2-y^2)}{\rm
Im}(\bar\Phi_{,x}\Phi_{,y}). \label{S_xy} \ee

The substitution of (\ref{EF_msm}) and (\ref{metric_msm}) into
(\ref{S_xy}) yields the following result: \be
S^\varphi=\frac{qb\kappa^5(x^2-y^2)^3 F(x,y)}{64\pi
\sqrt{E}D^{5/2}(m^2-q^2)^6}, \label{S_msm} \ee where $F(x,y)$ in
the numerator of $S^\varphi$ is some function of the coordinates
$x$ and $y$ which is not given here explicitly because of its
complicated form. However, in order the reader could have an idea
of how the function $F(x,y)$ looks like, below we give its
expression in the equatorial plane ($y=0$): \bea F(x,y=0)&=&\{
x(\kappa x+m)[(m^2-q^2)^2+4b^2]+\kappa(m^4-q^4)\} \nonumber\\
&&\times[(m^2-q^2)^2(1-x^2)-4b^2x^2]^5. \label{F_y0} \eea

In the above example the numerator of $S^\varphi$, as expected,
contains the factor $qb$, and vanishing of either the electric
charge $q$ or magnetic dipole moment $b$ causes $S^\varphi$ to
vanish too. When both parameters $q$ and $b$ have non--zero
values, the frame--dragging effect takes place which gives birth
to the flow of electromagnetic energy in the $\varphi$--direction
predicted by Bonnor.

%%%%%%%%%%%%%%%%%%%%%%
\section{Conclusion}
%%%%%%%%%%%%%%%%%%%%%%

In this paper we have succeeded in demonstrating that the use of
Ernst's complex potentials formalism simplifies considerably the
study of Poynting vector in stationary electrovacuum spacetimes
with axial symmetry. The formulae obtained by us for the only
non--zero component of Poynting vector require exclusively the
knowledge of the electromagnetic Ernst potential $\Phi$ for
establishing the presence or absence of the azimuthal
electromagnetic energy flows in a given spacetime. The component
$S^\varphi$ turns out to be invariant under the duality rotations
of the potential $\Phi$, which helps to single out special cases
where the presence of both the electric and magnetic fields does
not produce the frame--dragging effect. By direct calculation we
have confirmed Bonnor's prediction that $S^\varphi$ does not
vanish in spacetimes representing a static mass endowed with both
the electric charge and magnetic dipole moment.

\ack

This work was partially supported by Project 45946--F from CONACyT
of Mexico.

\end{document}